Natalia Ożegalska-Łukasik[1]
Jagiellonian University in Krakow, Poland
ORCID: 0000-0002-4239-4429
Szymon Łukasik[2]
AGH University of Krakow, NASK – National Research Institute, Poland
ORCID: 0000-0001-6716-610X

# CULTURALLY RESPONSIVE ARTIFICIAL INTELLIGENCE – PROBLEMS, CHALLENGES AND SOLUTIONS

Abstract

In the contemporary interconnected world, the concept of cultural responsibility occupies paramount importance. As the lines between nations become less distinct, it is incumbent upon individuals, communities, and institutions to assume the responsibility of safeguarding and valuing the landscape of diverse cultures that constitute our global society. This paper explores the socio-cultural and ethical challenges stemming from the implementation of AI algorithms and highlights the necessity for their culturally responsive development. It also offers recommendations on essential elements required to enhance AI systems' adaptability to meet the demands of contemporary multicultural societies. The paper highlights the need for further multidisciplinary research to create AI models that effectively address these challenges. It also advocates the significance of AI enculturation and underlines the importance of regulatory measures to promote cultural responsibility in AI systems.

**Introduction**

Cultural responsibility is one of the critical competencies in the modern world. As borders blur and the interconnectedness of nations intensifies, it is imperative that individuals, communities, and institutions take responsibility for preserving and respecting the diverse cultural tapestry that makes up our global society.

Researchers in the area of inclusiveness use three terms to describe the ability to adapt to multicultural settings. Cultural competence entails being able to function within the context of the cultural beliefs, behaviours, and needs of people and their communities (Thorpe and Williams-York, 2012). Cultural responsiveness emphasises listening to people's understanding and their experiences and adapting one's behaviour to their preferences (Miller et al., 2019). Cultural humility refers to the process of self-reflection and understanding one's own implicit and explicit biases and how these biases may influence behavioural patterns (Miller et al., 2019).

---

[1] natalia.ozegalska@uj.edu.pl
[2] slukasik@agh.edu.pl

The importance of cultural responsibility in education, business, health care, and legal spheres has already been highlighted in the respective fields, and there is a significant body of literature studying various aspects of its implementation in these areas. At the same time, while engineers are increasingly expected to work effectively across countries and cultures, their traditional trade was defined almost entirely by their technical skills. This has been rapidly changing in recent years, in Information Technology in particular. The ACM (Association for Computing Machinery) Special Interest Group on Computer Science Education in 2001 stressed that "cultural dimensions of Information Technology (IT) can no longer be ignored, with the expansion of the global economy, global markets and global communication enabled by information technology" (Little et al., 2001). Today, it is necessary for IT engineers to develop those other skills and cross-curricular competences that contribute to the overall development of their professions, both at school and in the workplace (Rico-Garcia, Fielden Burns, 2020). This is especially true for the omnipresent Artificial Intelligence (AI) technologies surrounding us, influencing not only the way we live but also the way we perceive the world around us.

The aim of this paper is to discuss the socio-cultural and ethical problems arising from the implementation of AI algorithms. We highlight the challenges this domain must address with regard to its culturally responsive development. Furthermore, we offer a discussion and recommendations regarding the essential elements necessary to enhance the adaptation of AI systems to meet the demands of contemporary multicultural societies.

**The concept of multiculturalism and its importance**

Our world is characterised by an exceptional degree of cultural, social, and religious diversity. Multiculturalism is an undeniable and dominant feature of the reality in which we function. The discourse around it has arisen in connection with the rapid process of globalisation, migration, and the opening of borders, and as a consequence of conflicts. However, the analysis of this phenomenon requires an interdisciplinary approach, given that its effects cover all aspects of life.

Cultural diversity has been known since ancient times, although in the Middle Ages, the dominant vision of the world tried to see it in the most homogeneous form, including a simple division of "us versus them". Moreover, until the end of the Second World War, as Will Kymlicka (2010) noted, in the Western world, relations between groups were regulated by an ideology clearly proclaiming the superiority of one culture over another, and thus granting the right of one group to subordinate another. Relations between ethno-cultural groups could be described using dichotomies: conquerors – conquered; colonialists – colonised; masters – slaves; settlers – natives; civilised – primitive; and orthodox – heretics, etc. (Kwiatkowska, 2019).

John Berry and David Sam, examining the Canadian context, introduced the concept of multiculturalism for the first time, dividing it into: a) demographic reality; b) ideological position; c) government policy; and d) a philosophical concept developed by political philosophers (Kymlicka, 2010). According to Golka (2010), multiculturalism means the coexistence in the same geographical area of two or more groups that differ in appearance, language, religion, or value system. These disproportions lead to different ways of perceiving reality (Golka, 2010). In turn, Nikitorowicz (2015) characterises multiculturalism as "a complex, dynamic process occurring in the context of the presence and mutual experience of racial, ethnic, religious, linguistic, national and traditional differences in a specific space". Paleczny (2019) emphasises that multiculturalism is a conscious, ordered state of cultural diversity.

There are many variants and varieties of multiculturalism, depending on the number of components, their size, the reasons for their creation, and their presence in the multicultural space. Multiculturalism, structured structurally, normatively, and politically, creates a specific model of a pluralistic society. Alfred Kroeber (2007) rightly notes that "every cultural phenomenon must be understood and assessed in the context of the culture to which it belongs." In this context, understanding culture is crucial because it serves as a fundamental concept regarding multiculturalism. It is the coexistence of various cultures that creates unique mosaic compositions.

Multiculturalism, cultural diversity, and interculturalism are all concepts denoting phenomena, the scale of which is becoming a sign of modernity and has deep and lasting consequences for both individuals and entire social groups. However, many of these phenomena have long been taking place beyond the control of political institutions and organisations and are moving into cyberspace, accompanying the processes of creating new types of bonds in the "network" society (Castells, 2008). The technological revolution has created a new communication space in which cultural responsibility is blurred. People and the communities they create increasingly go beyond the clear, axiological, normative, including linguistic, religious, and ethical boundaries of traditional ethnic, national, and civilisational cultures.

The phenomena of going beyond known symbolic areas, called transculturation or transgression, is accompanied by social crises and identity, a phenomenon which with, as yet, Artificial Intelligence is unable to cope in sufficient manner. The visible crisis of multiculturalism policy in the world is leading to many consequences. Multiculturalism policy, in accordance with ideological and theoretical assumptions, should lead to the construction of legal and social spaces for the integration of diverse cultural groups in an AI-based reality. To present the stage we are currently at while implementing the elements of cultural responsibility in the field of AI, first, we will formally introduce this domain in more detail.

**Artificial intelligence – concept and ethical background**

The term Artificial Intelligence is widely attributed to John McCarthy of the Massachusetts Institute of Technology, who coined this term while trying to secure funding for a summer research project in 1955 (Neri & Cordeiro, 2020). At that time, McCarthy defined the goal of AI as being to "write a calculator program which can solve intellectual problems as well as or better than a human being in areas like program writing, theorem proving or gameplay" (Penn, 2021). Currently, after almost 70 years of development in theories and practice and encountering several crises (so-called "winters of AI"), Artificial Intelligence can be understood as "a system's ability to interpret external data correctly, to learn from such data, and to use those learnings to achieve specific goals and tasks through flexible adaptation" (Haenlein & Kaplan, 2019).

The landscape of AI methods includes a variety of techniques based on inspiration from nature. In recent years, this field has predominantly used Artificial Neural Networks (ANN), which are based on simplified models of neuronal cells. The creation of models spanning across multiple layers with appropriate learning methods has led to the emergence of the Deep Learning field, which allows the building of complex data-driven algorithms (Sarker, 2021). Deep Learning models can operate as discriminators or generators. The first of these construct decision boundaries in data space – an ability which is particularly useful in a classification task where the algorithm's goal is to assign data elements to one of the predefined classes presented (e.g., recognising the writer of a text, an object in an image, or the author of a song). Generative models are trained on large datasets to learn the data's patterns, structures, and features. They then use this knowledge to generate new data that is consistent with what they have learned. In recent years, this has led to a plethora of applications, including art, content creation, data augmentation, and simulation (Ramdurai &

Adhithya, 2023). While most AI research concentrates on ANNs, inspiration stemming from nature can be found in other techniques. Among others, one should name evolutionary algorithms – optimisation and search techniques inspired by the process of natural evolution, swarm intelligence – inspired by the collective behaviour of social organisms, such as ants or bees, or fuzzy logic – a mathematical framework that deals with uncertainty and imprecise information (Al Mansoori et al., 2019).

The development of Artificial Intelligence algorithms has sparked a discussion on their ethical implications. This can be traced back to the pioneers of AI and the early days of information technology in general. In 1942, in his short story "Runaround", the science fiction writer Isaac Asimov introduced the Three Laws of Robotics, namely:

1. *A robot may not injure a human being or, through inaction, allow a human being to come to harm.*
2. *A robot must obey the orders given it by human beings except where such orders would conflict with the First Law.*
3. *A robot must protect its own existence as long as such protection does not conflict with the First or Second Law.* (Asimov, 1942)

The revolutionary development of Artificial Intelligence algorithms in recent years has stimulated the creation of new regulations that strictly define the legal and ethical framework for the use of AI techniques, generative AI in particular. These include the "AI Act" developed within the organisational structures of the European Union (European Parliament, 2023), "Interim Measures for Generative Artificial Intelligence Service Management" passed recently in the People's Republic of China (Cyberspace Administration of China, 2023) or an "AI Bill of Rights – Blueprint" (White House, 2023). Most of them indirectly try to address problems related to adapting AI algorithms to the multicultural reality – an aspect which will be discussed in the next section.

**Culturally responsive AI – current landscape**

The established notion of digital humanism, as initially put forth by Doueihi (2011), has allowed us to transcend the conventional stereotype of conflict between humans and machines. Expanding upon this idea, Sinatra and Vitali-Rosati (2014) take it a step further by asserting that "digital technology is, in fact, an entity that actively shapes culture, engendering a fresh perspective on the world and giving rise to a novel 'civilization'". As pointed out by Adams:

> *Advances in Artificial Intelligence (AI) will have subtle effects on individuals and on culture. They may create new knowledge, make certain types of knowledge more accessible, and change the value of some types of knowledge and ways of thinking. Even though the exact form of these effects is unpredictable, AI researchers have an ethical responsibility to evaluate their work from this perspective* (Adams, 1986)

One can therefore categorically state that this responsibility includes the inclusion of multicultural aspects in the design and application of AI systems. The realisation of culturally responsible AI algorithms faces many challenges arising both from their method of operation and implementation.

One of the significant issues discussed within the AI community is the problem of cultural bias. Bias is understood as the inclination or prejudice of a decision made by an AI system that is for or against one person or group, especially in a way considered to be unfair (Ntoutsi, 2020). Typically, bias enters the AI system through the data comprising the input to the given algorithm (so-called training data bias). Algorithmic bias, on the other hand, is introduced through the flawed design of the algorithm itself, causing its outputs to benefit or disadvantage specific individuals or groups more than others without a justified reason for such unequal impacts (Kordzadeh & Ghasemaghaei, 2022). Artificial Intelligence systems are known to amplify existing cultural biases. This demonstrates itself in generative AI (see Figure 1 for an example) but also in discriminative use cases. Another well-known example is a racial bias found by Obermeyer et al. (2019) in the commercial algorithm used to guide health decisions for the US healthcare system. The authors estimated that this racial bias reduces the number of patients who self-identified themselves as black assigned for extra care by more than half. In another experiment, Algorithm Watch – a human rights organisation investigating the role of algorithms in the modern world – showed that Google Vision Cloud, a computer vision service, labelled an image of a dark-skinned individual holding a thermometer "gun" while a similar image with a light-skinned individual was labelled

"electronic device". A subsequent experiment showed that the image of a dark-skinned hand holding a thermometer was labelled "gun" and that the same image with a salmon-coloured overlay on the hand was enough for the computer to label it "monocular" (Kayser-Bril, 2020). While similar cases are circulating on the Internet, as discovered by inquisitive users, one has to note additional side-effects of cultural bias. Recent AI algorithms can analyse patterns and styles from different works of art and culture and are able to mimic these patterns. Even the abundance of available input data does not eliminate the risk of reducing the diversity of the resulting products and both their authenticity and cultural significance. Taking it to the extreme level can lead to cultural hegemony. In a recent study, Cao et al. (2023) discovered that OpenAI's ChatGPT exhibits a strong alignment with American culture but adapts less effectively to other cultural contexts. Furthermore, by using different prompts to probe the model, they have demonstrated the flattening of cultural differences and biasing them towards American culture.

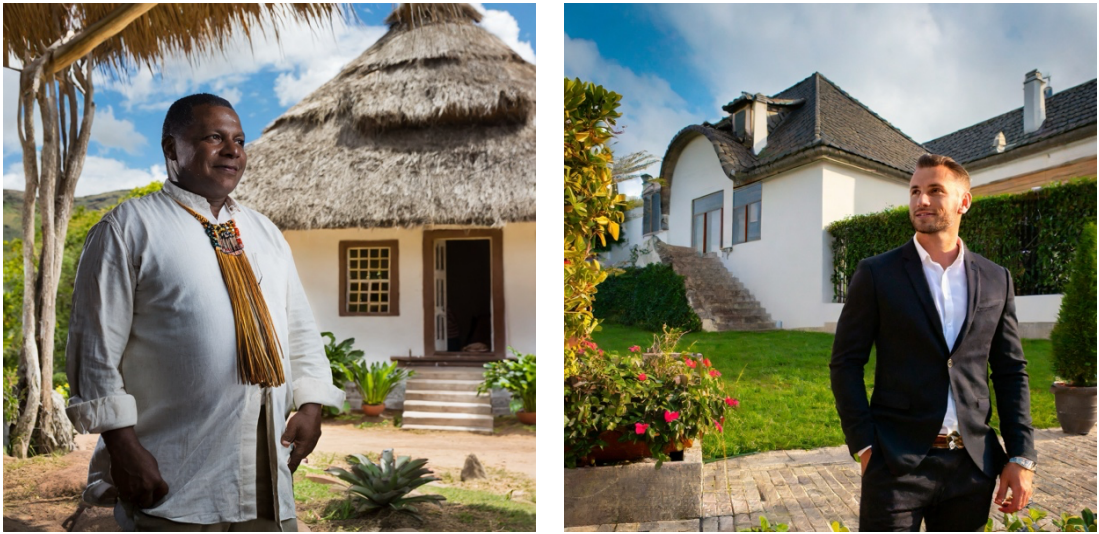

**Figure 1.** An example of AI amplifying cultural bias: the generation of the image "Wealthy African man and his house" vs. "Wealthy European man and his house" / authors' own work using proprietary Generative AI algorithm /.

The other important issue is that AI algorithms can contribute to technological exclusion across different cultures. Despite global progress in extending the use of the Internet and Information Communications Technology (ICT), it is the countries with well-developed R&D sectors and the possibility to collect and process data that can engage greater resources for the development of AI algorithms adapted to the local perspective. Consequently, even widely available tools such as ChatGPT do not demonstrate similar performance across different cultural areas. With a 20% difference in the performance between benchmarks in English and Telugu (as seen in Figure 2) – a language spoken in India by about 96 million people (Eberhard et al., 2023) – GPT-4 model (powering the newest implementations of ChatGPT) cannot be seen as a universal tool not favouring English-speaking users. Existing biometric systems – primarily due to training data biases – also tend to exhibit similar tendencies. Researchers affiliated with the MIT Gender Shades project discovered that few commercial face-recognition algorithms underperform in the task of gender classification when used for females with dark-coloured skin (see Fig. 3 for results of this study).

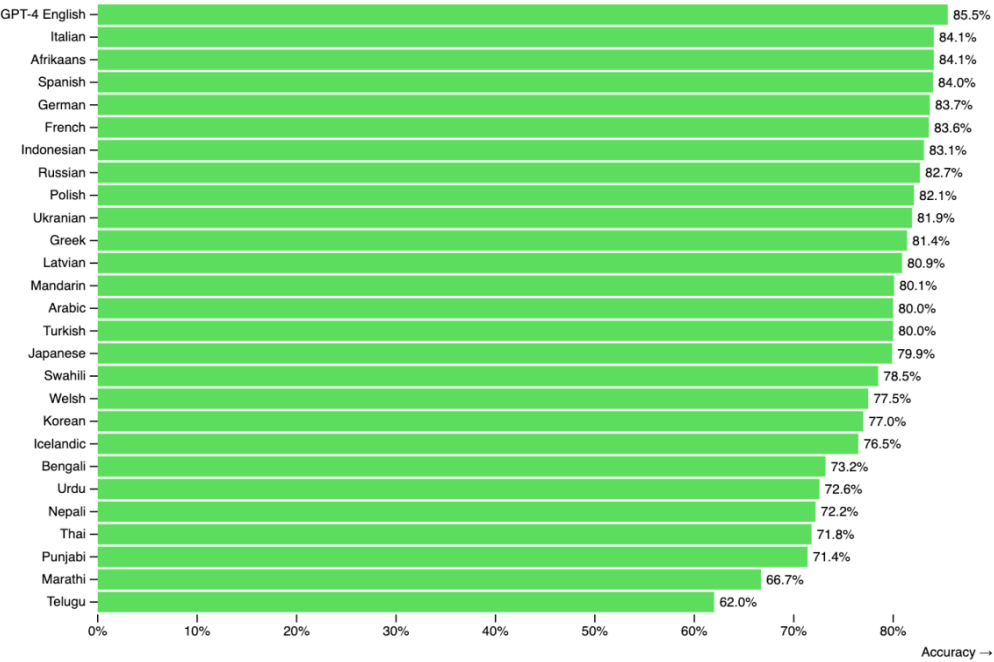

**Figure 2.** An example of technological exclusion across cultures – the performance of GPT-4 model across different languages on an MMLU AI benchmark – a suite of 14,000 multiple-choice problems spanning 57 subjects /source: (OpenAI, 2023)/.

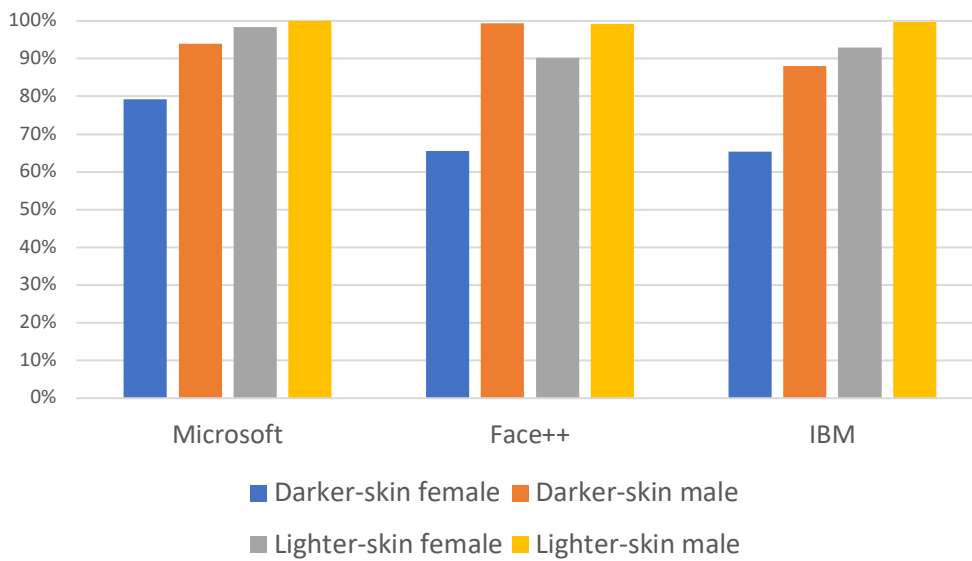

**Figure 3.** Accuracy of three commercial face-recognition algorithms in gender classification for varying skin colours /authors' own work based on (Buolamwini & Gebru, 2018)/.

Serious adverse effects were also identified in AI-based clinical models – with examples such as varying sensitivity and specificity between race/ethnicity groups in dementia status classification, differences in error rates in ICU mortality between racial groups, or lower diagnostic accuracy in darker-skinned individuals compared to lighter-skinned individuals for diagnosis of diabetic retinopathy (Huang et al., 2022).

The overview presented above does not cover all the issues related to the cultural responsibility of the AI ecosystem. Less direct elements, such as the unequal economic impact of widespread implementation of algorithms and its effect on societies, are also part of the bigger picture. Responding to these challenges requires actions from numerous actors involved – not only the designers of the algorithm themselves but also governing bodies and end-users. Possible solutions and recommendations will be covered in the next section.

**Recommendations**

The need for further research on AI systems, particularly from a multicultural and ethical perspective, is paramount in today's rapidly evolving technological landscape. Recent research on AI, notably generative systems such as GPT-4, has brought novel ethical and cultural challenges to the forefront. These systems have demonstrated unprecedented capabilities in generating text, images, and even code, and consequently, they have also introduced a transformative dimension to human-computer interactions. Artificial intelligence combined with human ingenuity gives an incredible range of possibilities to solve problems in a highly creative way. However, the real challenge in the context of using artificial intelligence is discovering cases and identifying complex, interdisciplinary problems that can be solved using it. Finding problems, as opposed to solving them, requires cognitive diversity – which tends to be suppressed in homogeneous cultures (Kumar, 2023). Hence, individuals engaged in research and education who possess the ability to pose insightful inquiries, venture beyond conventional paradigms, grasp the broader context, articulate the "what" and "why" of occurrences, and adopt a perspective that embraces diverse viewpoints play a vital role in propelling the field of Artificial Intelligence forward.

From the engineering perspective, there has been a lot of effort in the research and examination of AI systems' performance in terms of cultural responsibility. Generative AI systems undergo a rigorous testing procedure while also being constantly monitored using both user reviews and automatic (also partly AI-based) systems. Persevering researchers detect biases by forcing generative algorithms to create code fragments whose output is intended to highlight the unequal treatment of specific cultural factors (Biddle, 2022). Up to this point, however, this has been primarily an independent effort or the responsibility of the algorithm's designer, not normalised in any way. While there is no uniform stance on this issue, many AI experts and organisations support the idea of AI regulation and the establishment of formal standards. They argue that regulations are essential to address safety, accountability, transparency, and fairness concerns in AI systems. They also believe that clear guidelines can help prevent misuse and negative consequences. The European proposal of the AI Act takes the risk-based approach – with the responsibility for risk assessment lying with the economic operators placing AI systems on the market. While the Act does not directly refer to the importance of cultural responsibility and evaluating the cultural impact of AI algorithms, it underlines the significance of detecting, monitoring and correcting biases in order to protect the rights of others from discrimination (European Commission, 2023). This law – which is perceived as one of the first comprehensive attempts to regulate AI implementation – offers no distinct guidelines for classifying AI systems, carrying out requirements, or conducting assessments. Thus, it is up to the Member States, stakeholders, and experts in this domain to outline the essential details that will have a critical impact on the issues discussed here.

Finally, the importance of AI enculturation must be underlined. This concept was derived from machine enculturation which is a method employed to ensure that sociocultural values are present in computers in order that they can more readily relate to humans and avoid social disruptions or even physical harm (Messner, 2022). It has been recently demonstrated that, for instance, altering the "brushstroke of cultural features" that make objects perceived as belonging to a given culture while preserving their functionalities is possible (Zaino et al., 2022). Including cultural factors can also improve the performance of discriminative AI algorithms – which is also encouraging results for practitioners seeing a need to take these aspects into account (Messner, 2022)

**Conclusion**

The global discussion on Artificial Intelligence seems to be organised in circular processes coupled with newly appearing engineering solutions. We start with a general fascination with possibilities (what Artificial Intelligence can do), then move on to a sense of unreflective optimism (Artificial Intelligence will save the world) and reach a moment of overwhelming fatalism and pessimism (Artificial Intelligence will destroy the world). It cannot be denied that the development of Artificial Intelligence poses several serious challenges that we must face. Indeed, AI reflects the people who created it. Part of the human population shows tolerance and acceptance of other cultures and others – on the contrary – do not. Artificial Intelligence systems, if they are not subject to control and regulation (including legal standards), could exhibit similar tendencies. Humanity is therefore tasked with creating the next generations of AI models that will respond more effectively to the challenges of modern, multicultural societies. This is because in this globalised world, embracing cultural responsibility is not merely an option but an ethical imperative, as it holds the key to fostering a sense of inclusivity and tolerance that is essential today more than ever.